\documentclass {article}

\newcommand {\address} [1]
  {}
\newenvironment {references}
  {\begin {thebibliography} {99}}
  {\end {thebibliography}}

\newtheorem {lemma}{Lemma}
\newtheorem {prop}{Proposition}
\newtheorem {definition}{Definition}
\newtheorem {example}{Example}
\newtheorem {theorem}{Theorem}
\newenvironment {proof}
  {\par\parindent 0pt Proof\par}
  {\par\hfill$\bf\Omega$}
\newenvironment {proofprop}{\par\parindent 0pt Proof of the proposition\par}
  {\par\hfill$\bf\Omega$}

\newcommand\Tr {\,{\rm Tr}\,}

\begin {document}

\title {Rigorous results in non-extensive thermodynamics}
\author{Jan Naudts\\
\small
Departement Natuurkunde, Universiteit Antwerpen, UIA,
\\
\small
Universiteitsplein 1, 2610 Antwerpen, Belgium
}
\date {August 1999 v4}
\maketitle

\begin {abstract}
This paper studies quantum systems with a finite number of degrees of
freedom in the context of non-extensive thermodynamics. A trial density
matrix, obtained by heuristic methods, is proved to be the equilibrium
density matrix. If the entropic parameter $q$ is larger than 1 then
existence of the trial equilibrium density matrix requires that $q$ is
less than some critical value $q_c$  which depends on the rate by which
the eigenvalues of the hamiltonian diverge.  Existence of a unique
equilibrium density matrix is proved if in addition $q<2$ holds. For $q$
between 0 and 1, such that $2<q+q_c$, the free energy has at least one
minimum in the set of  trial density matrices. If a unique equilibrium
density matrix exists then it is necessarily one of the trial density
matrices. Note that this is  a finite rank operator, which means that in
equilibrium high energy levels have zero probability of occupancy.

\end {abstract}

\section {Introduction}

The formalism of non-extensive thermodynamics started more than 10 years
ago with the introduction by C. Tsallis \cite {TC88} of a family of
entropies, parameterized with a parameter $q$ called the {\sl entropic
parameter}. It has  developed gradually to a collection of mostly
phenomenological results with a large number of applications, some more
convincing than others. Nevertheless, part of the physics community is
still sceptic about the need and physical relevance of the theory. By
providing mathematical proofs for some of the fundaments of the 
formalism this paper tries to improve its credibility and to provide the
necessary base for further extension.

Some of the results presented in the present paper, in particular concerning
$q>1$-statistics, have been published already in \cite {NC99}.
Here, missing details are filled in and the $q<1$-case is added.

The structure of the paper is as follows. In the next section the
necessary concepts are introduced. In section 3 the main theorem about
$q>1$-statistics is formulated. Its proof follows in two consecutive
sections. Section 6 is the $q<1$-version of section 3. In section 7
is shown that the free energy has at least one minimum in the set of
trial density matrices. In section 8  the main result about
$q<1$-statistics is formulated and proved. Finally, section 9 gives a
short summary and discussion of the obtained results.

\section {Canonical ensemble}

This paper is limited to quantum mechanical systems with a finite number
of degrees of freedom. The state of such a system is described by a
density matrix $\rho$ (i.e., $\rho\ge 0$, trace-class and $\Tr\rho=1$)
on a finite dimensional or separable Hilbert space ${\cal H}$.

For any $q>0, q\not=1$, and for any density matrix $\rho$ of ${\cal H}$,
the Tsallis entropy is defined by
\begin {equation}
{\cal S}_q(\rho)=k_B{1-\Tr\rho^q\over q-1}
\end {equation}
If for small $q$ the operator $\rho^q$ is not trace-class then put
${\cal S}_q(\rho)=+\infty$. $k_B$ is Boltzmann's constant.
It is introduced for historical reasons. 
$q$ is called the {\sl entropic parameter}.
Note that in the limit of
$q=1$ the Shannon entropy is recovered. Indeed, one has
\begin {equation}
\lim_{q\downarrow 1}{\cal S}_q(\rho)=-k_B\Tr\rho\ln \rho
\label {shannon}
\end {equation}
Note also that always ${\cal S}_q(\rho)\ge 0$.
The equality ${\cal S}_q(\rho)=0$ implies that 
$\rho$ is the orthogonal projection onto a one-dimensional subspace
of ${\cal H}$.

The thermodynamic formalism is based on a pair consisting of entropy
together with energy. Energy is defined in terms of a hamiltonian
$H$ which is a self-adjoint operator of ${\cal H}$.
Throughout the paper it is 
assumed that $H$ has a discrete spectrum bounded from
below, and that all eigenvalues have finite multiplicity.
More precisely, there exists an orthonormal basis
$(\psi_n)_{n\ge 0}$ of ${\cal H}$ such that $H\psi_n=\epsilon_n\psi_n$
for all $n$, with eigenvalues $\epsilon_n\in{\bf R}$ ordered increasingly.
If $\cal H$ is finite dimensional then it is assumed that $H$ is
not a multiple of the identity.

For the existence of an equilibrium density matrix it is important that the
eigenvalues $\epsilon_n$ tend to infinity fast enough as
$n\rightarrow\infty$. Let us therefore introduce

\begin {definition}
The {\sl critical entropic parameter} $q_c$ of $H$ is the upper limit
of $q\ge 1$
for which constants $\alpha$ and $\gamma$ exist such that the sequence
\begin {equation}
\alpha+\epsilon_n\ge\gamma n^{q-1}, n=1,2,\cdots
\end {equation}
is bounded.
\end {definition}

\noindent
In the terminology of Connes \cite {CA94} the operator $(\alpha+H)^{-1}$,
$\alpha$ large enough, is an infinitesimal of order $q-1$
for all $q\in (1,q_c)$ -- see e.g.~\cite {LG97}, section 5.
If the operator $H$ is bounded then $q_c=1$. For the harmonic oscillator
$q_c=2$ holds, for a particle enclosed in a $d$-dimensional box is $q_c=1+2/d$.

For the energy ${\cal U}_q$
several propositions have been made\cite {TC88}\cite {CT91}, the latest
of which is \cite {TMP98}
\begin {equation}
{\cal U}_q(\rho)={\Tr\rho^q H\over \Tr\rho^q}
\equiv{1\over \Tr\rho^q}\sum_{n=0}^\infty\epsilon_n(\rho^q\psi_n,\psi_n)
\le+\infty
\label {energ}
\end {equation}

The density matrix $\rho$ is an
equilibrium density matrix at temperature $T>0$
if the free energy ${\cal F}^\beta$, given by
\begin {equation}
{\cal F}^\beta_q(\rho)={\cal U}_q(\rho)-T{\cal S}_q(\rho)
\label {freeenerg}
\end {equation}
has a unique minimum at $\rho$ (as usual $\beta=1/k_BT$, with $k_B$
Boltzmann's constant).

The remainder of the paper is concerned with the existence of equilibrium 
states. The two cases $q>1$ and $q<1$ behave differently. The case $q>1$ is 
considered first.

\section {$q>1$-statistics}

Heuristic arguments lead to the conclusion that the equilibrium density
matrix, if it exists,
is of the form $\rho_\alpha$ given by
\begin {equation}
\rho_\alpha={1\over\zeta_\alpha}
\left({1\over \alpha{\bf 1}+ H}\right)^{1/(q-1)}
\end {equation}
and
\begin {equation}
\zeta_\alpha=\Tr\left({1\over \alpha{\bf 1}+ H}\right)^{1/(q-1)}
\end {equation}
The parameter $\alpha$ should satisfy $\alpha >-\epsilon_0$ to
guarantee that the denominator $\alpha{\bf 1}+H$ is strictly positive.
If ${\cal H}$ is infinite dimensional then
the entropic index $q$ should satisfy $q<q_c$ to ensure
that $\rho_\alpha$ is a trace-class operator.

\begin {prop}
If ${\cal H}$ is finite
dimensional or if $1<q<q_c$ then the energy ${\cal U}_q(\rho_\alpha)$
is well-defined and finite for all $\alpha>-\epsilon_0$. One has
\begin {equation}
{\cal U}_q(\rho_\alpha)
={1\over \zeta_\alpha^{q-1}}{1\over\Tr\rho_\alpha^q}-\alpha
\label {energlo}
\end {equation}
\end {prop}

\begin {proof}
If ${\cal H}$ is finite dimensional the statement is obvious.
So let us assume that ${\cal H}$ is infinite dimensional.

Because $q>1$ and $\rho_\alpha$ is trace-class
one has automatically that also $\rho_\alpha^q$
is trace-class. Note that
\begin {eqnarray}
\rho^q_\alpha H
&=&{1\over\zeta_\alpha^{q-1}}
{\rho_\alpha H\over \alpha{\bf 1}+ H}\cr
&=&{1\over \zeta_\alpha^{q-1}}
\rho_\alpha\left({\bf 1} - {\alpha{\bf 1}\over {\bf 1}+\alpha H}\right)\cr
&=&{1\over \zeta_\alpha^{q-1}}
\rho_\alpha - \alpha\rho_\alpha^q
\label {energrel}
\end {eqnarray}
This implies (\ref {energlo}). Since both $\rho_\alpha$
and $\rho_\alpha^q$ are trace-class
one has ${\cal U}_q(\rho_\alpha)<+\infty$.
\end {proof}

From (\ref {energlo}) follows
that the free energy equals
\begin {equation}
{\cal F}^\beta_q(\rho_\alpha)=
{1\over \zeta_\alpha^{q-1}}{1\over\Tr\rho_\alpha^q}-\alpha
-{1\over\beta(q-1)}(1-\Tr\rho_\alpha^q)
\end {equation}
Variation w.r.t.~$\alpha$ gives
\begin {equation}
{\partial\ \over\partial\alpha}
{\cal F}^\beta_q(\rho_\alpha)=
-(q-1){1\over\zeta_\alpha^q}{1\over\Tr\rho_\alpha^q}
{\partial\zeta_\alpha \over\partial\alpha}
-1
-\left({1\over\zeta_\alpha^{q-1}}{1\over(\Tr\rho_\alpha^q)^2}
-{1\over\beta(q-1)}\right){\partial\ \over\partial\alpha}\Tr\rho_\alpha^q
\end {equation}
Using
\begin {equation}
{\partial\ \over\partial\alpha}\zeta_\alpha=-{1\over q-1}
\Tr{1\over (\alpha{\bf 1}+H)^{q/(q-1)}}
=-{1\over q-1}\zeta_\alpha^q\Tr\rho_\alpha^q
\label {dzetadalpha}
\end {equation}
there follows
\begin {equation}
{\partial\ \over\partial\alpha}{\cal F}^\beta_q(\rho_\alpha)
={1\over q-1}
\left({1\over\beta}-{1\over\beta_q(\alpha)}\right)
{\partial\ \over\partial\alpha}\Tr\rho_\alpha^q
\label {fcond}
\end {equation}
with
\begin {equation}
\beta_q(\alpha)
={1\over q-1}\zeta_\alpha^{q-1}(\Tr\rho_\alpha^q)^2
\end {equation}

\begin {prop}
\label {strictdecr}
The function
$\Tr\rho_\alpha^q$ is strictly decreasing in $\alpha$.
\end {prop}

\begin {proof}
A short calculation using (\ref {dzetadalpha}) gives
\begin {eqnarray}
(q-1){\partial\ \over\partial\alpha}\Tr\rho_\alpha^q
&=&(q-1){\partial\ \over\partial\alpha}
\zeta_\alpha^{-q}\Tr(\alpha{\bf 1}+H)^{q/(1-q)}\cr
&=&-q(q-1)\zeta_\alpha^{-q-1}\Tr(\alpha{\bf 1}+H)^{q/(1-q)}
{\partial\zeta_\alpha \over\partial\alpha}\cr
& &\qquad
-q\zeta_\alpha^{-q}\Tr(\alpha{\bf 1}+H)^{(2q-1)/(1-q)}\cr
&=&q(\zeta_\alpha^{-q-1}\left(\Tr(\alpha{\bf 1}+H)^{q/(1-q)}\right)^2\cr
& &\qquad
-q\zeta_\alpha^{-q}\Tr(\alpha{\bf 1}+H)^{(2q-1)/(1-q)}\cr
&=&q\zeta_\alpha^{-q-1}\bigg[f_\alpha(q)^2-f_\alpha(1)f_\alpha(2q-1)\bigg]
\label {eqa}
\end {eqnarray}
with
\begin {equation}
f_\alpha(x)=\Tr (\alpha{\bf 1}+H)^{x/(1-q)}
\label {falfdef}
\end {equation}
Now, the function $f_\alpha$ is strictly log-convex because $H$ is not a
multiple of the identity (see the appendix).
Hence the r.h.s.~of (\ref {eqa}) is negative.
This ends the proof of the proposition.
\end {proof}

From proposition \ref {strictdecr} follows that (\ref {fcond})
can vanish only if
\begin {equation}
\beta=\beta_q(\alpha)
\label {tcond}
\end {equation}
Note that $\beta_q(\alpha)$ can be written out as
\begin {equation}
\beta_q(\alpha)={1\over q-1}
{\left(\Tr(\alpha{\bf 1}+ H)^{-q/(q-1)}\right)^2
\over \left(\Tr(\alpha{\bf 1}+ H)^{-1/(q-1)}\right)^{1+q}
},
\qquad \alpha>-\epsilon_0
\label {betaexplic}
\end {equation}

\begin {prop}
\label {uniqsolprop}
$\beta_q$ is a strictly decreasing function
of $\alpha>-\epsilon_0$, with range $(0,+\infty)$.
\end {prop}

\begin {proof}
Take the logarithm of $(q-1)\beta_q(\alpha)$.
Its derivative w.r.t.~$\alpha$ equals
\begin {equation}
-{2q\over q-1}{f_\alpha(2q-1)\over
f_\alpha(q)}+{1+q\over q-1}{f_\alpha(q)\over f_\alpha(1)}
\label {logrhs}
\end {equation}
with $f_\alpha$ given by (\ref {falfdef}).
Since $f_\alpha$ is strictly log-convex, and $q+1<2q$,
and $q=(1/2)1+(1/2)(2q-1)$,
expression (\ref {logrhs}) is negative. This implies that
$\beta_q$ is strictly decreasing in $\alpha$.

Recall that $\epsilon_0$ denotes the ground state
energy, i.e.~the lowest eigenvalue of $H$ and that
$\alpha>-\epsilon_0$ is required for the existence of $\rho_\alpha$.
In the limit $\alpha\downarrow -\epsilon_0$ the
function $\alpha\rightarrow f_\alpha(x)$ diverges as
\begin {equation}
m(\alpha +\epsilon_0)^{x/(1-q)}
\end {equation}
with $m$ the multiplicity of the eigenvalue $\epsilon_0$.
In this limit $\beta_q(\alpha)$
behaves as $m^{1-q}/(\alpha+\epsilon_0)$ which tends to $\infty$
as $\alpha$ tends to $-\epsilon_0$.

On the other hand, if $\alpha$ is large enough,
then
$\displaystyle\Tr(\alpha{\bf 1}+H)^{q/(1-q)}$
is less than 1 so that
\begin {equation}
\beta_q(\alpha)
\le {1\over q-1}\left({
\Tr(\alpha{\bf 1}+H)^{1/(1-q)}(\alpha{\bf 1}+H)^{-1}
\over \Tr(\alpha{\bf 1}+H)^{1/(1-q)}}\right)^{1+q}
\end {equation}
which tends to zero as $\alpha\uparrow\infty$,
because $(\alpha{\bf 1}+H)^{-1}$ tends to zero in norm.
Hence  takes on all values between 0 and $+\infty$.
\end {proof}

Let $\alpha_q(\beta)$ denote the inverse of the function $\beta_q(\alpha)$.
It is strictly decreasing on the domain $(-\epsilon_0,+\infty)$.

\begin {prop}
\label {uniqsolprop0}
The function
$\alpha\rightarrow {\cal F}_q^\beta(\rho_\alpha)$ defined on
$(-\epsilon_0,+\infty)$
has a unique minimum at $\alpha=\alpha_q(\beta)$.
\end {prop}

\begin {proof}
That ${\cal F}_q^\beta(\rho_\alpha)$ has a unique minimum at
$\alpha=\alpha_q(\beta)$ follows because this function is strictly
decreasing for $\alpha<\alpha_q(\beta)$ and strictly increasing
for $\alpha>\alpha_q(\beta)$, as can be seen from (\ref {fcond}).
\end {proof}

The previous propositions support the formulation of the following result.

\begin {theorem} 
\label {theoremsol}
Let $1<q\le 2$. Let $H$ be a self-adjoint operator of ${\cal H}$.
Assume that either
\begin {itemize}
\item ${\cal H}$ is finite dimensional and $H$ is not a multiple of
$\bf 1$
\end {itemize}
\noindent
or
\begin {itemize}
\item ${\cal H}$ is infinite dimensional,
the spectrum of $H$ is discrete, bounded from below,
with isolated eigenvalues of finite multiplicity,
and $q<q_c$.
\end  {itemize}
Then for all $\beta>0$ the free energy
${\cal F}^\beta_q(\rho)$ has a unique minimum.
It occurs at $\rho=\rho_\alpha$ with $\alpha=\alpha_q(\beta)$.
\end {theorem}

\noindent
The proof of this theorem follows later on.

The value of the free energy, energy, and entropy for the equilibrium
state are denoted $F(T)$,  $U(T)$ and $S(T)$ respectively
(i.e.~$F(T)={\cal F}^\beta_q(\rho_\alpha)$ with
$\alpha=\alpha_q(\beta)$, and so on). These quantities satisfy the
following thermodynamic relations.

\begin {prop}
Under the conditions of theorem \ref {theoremsol} is
\begin {equation}
{\hbox{d }\over\hbox {d}T}F(T)=-S(T)
\qquad\hbox{ and }
{\hbox{d }\over\hbox {d}T}U(T)> 0
\end {equation}
\end {prop}

\begin {proof}
The first expression follows from (\ref {freeenerg}) because in equilibrium
$\partial {\cal F}^\beta_q(\rho_\alpha)/\partial\alpha=0$. The monotonicity
of the energy $U(T)$ as a function of temperature is shown as follows.
One obtains from (\ref {energlo}), (\ref {dzetadalpha})
and proposition \ref {strictdecr},
\begin {equation}
{\partial\ \over\partial\alpha}{\cal U}_q(\rho_\alpha)
=-{1\over\zeta_\alpha^{q-1}}{1\over(\Tr\rho_\alpha^q)^2}
{\partial\ \over\partial\alpha}\Tr\rho_\alpha^q>0
\end {equation}
From (\ref {tcond}) follows
\begin {equation}
{\hbox {d}\beta\over\hbox {d}\alpha}
=\zeta_\alpha^{q-2}(\Tr\rho_\alpha^q)^2
{\hbox {d }\over\hbox {d}\alpha}\zeta_\alpha
+{2\over q-1}\zeta_\alpha^{q-1}(\Tr\rho_\alpha^q)
{\hbox {d }\over\hbox {d}\alpha}\Tr\rho_\alpha^q
\end {equation}
Using (\ref {dzetadalpha}) this simplifies to
\begin {equation}
{\hbox {d}\beta\over\hbox {d}\alpha}
={1\over q-1}\zeta_\alpha^{q-1}(\Tr\rho_\alpha^q)
\left[-\zeta_\alpha^{q-1}(\Tr\rho_\alpha^q)^2
+2{\hbox {d }\over\hbox {d}\alpha}\Tr\rho_\alpha^q
\right]
\end {equation}
Next use (\ref {tcond}) to obtain
\begin {equation}
{\hbox {d}\beta\over\hbox {d}\alpha}
={1\over q-1}\zeta_\alpha^{q-1}(\Tr\rho_\alpha^q)
\left[2{\hbox {d }\over\hbox {d}\alpha}\Tr\rho_\alpha^q
-\beta(q-1)
\right]
\end {equation}
The latter expression is strictly negative
--- see proposition \ref {strictdecr}.
The desired result $\hbox{d}U(T)/\hbox{d}T>0$
follows now by application of the chain rule.
\end {proof}

\section {Convexity arguments}

The origin of the variational principle, stating that the free energy
is minimal in equilibrium, is that entropy
${\cal S}_q(\rho)$ should be maximal under the constraint that the
energy ${\cal U}_q(\rho)$ has a given value. To study convexity
properties it appears to be easier to consider the equivalent
problem of minimizing
${\cal U}_q(\rho)$ under the constraint that the entropy
${\cal S}_q(\rho)$ has a given value. The reason for this is that
at constant energy the denominator of (\ref {energ}) is also constant.
By the method of Lagrange multipliers, the minimum of
\begin {equation}
{\cal G}^\beta_q(\rho)=\Tr\rho^q H-T{\cal S}_q(\rho)
\label {oldfreenerg}
\end {equation}
is the solution of the problem of minimizing
${\cal U}_q(\rho)$ given ${\cal S}_q(\rho)$. By a proper choice of the value
of ${\cal S}_q(\rho)$ one then obtains a solution of the original
variational principle. In this way theorem \ref {theoremsol} can be
proved. In what follows the above reasoning is worked
out in a rigorous manner.

Let be given an $\alpha\in{\bf R}$ for which $\alpha{\bf 1}+ H$
is strictly positive
(i.e.~$\alpha+\epsilon_0>0$ with $\epsilon_0$ the lowest eigenvalue
of $H$).
Introduce a norm $||\cdot||_\alpha$
on the bounded operators of $\cal H$ by
\begin {eqnarray}
||A||_\alpha^2
&=&\Tr(\alpha{\bf 1}+H)|A|^2\cr
&\equiv&\sum_n(\alpha+\epsilon_n)||A\psi_n||^2\le+\infty
\end {eqnarray}

\begin {prop}
\label {convex}
Let $1<q\le 2$ and $q<q_c$. Let $\alpha=k_BT/(q-1)$
and assume that $\alpha{\bf 1}+H>0$.
Then one has
\begin {equation}
{\cal G}^\beta_q(\rho)-{\cal G}^\beta_q(\rho_\alpha)
\ge {1\over 2}q(q-1)||\rho-\rho_\alpha||_\alpha^2
\label {underbound}
\end {equation}
for any density matrix $\rho$.
\end {prop}

\noindent
Note that in the present section the temperature $T$ can be negative.
The proof of the proposition is based on Klein's inequality (see
e.g.~\cite {RD69}, 2.5.2) which can be formulated as follows.

\begin {lemma}
\label {klein}
Let $A$ and $B$ be self-adjoint operators with discrete
spectrum. Assume $B$ is diagonal in the basis $(\psi_n)_n$
of eigenvectors of $H$. Assume $\alpha{\bf 1}+H\ge 0$.
Then for any convex function $f$ one has
\begin {equation}
\Tr(\alpha{\bf 1}+H)(f(A)-f(B)-(A-B)f'(B))\ge 0
\end {equation}
\end {lemma}

\begin {proof}
Let $(\phi_n)_n$ be an orthonormal basis in which $A$ is diagonal.
I.e., $A\phi_n=a_n\phi_n$ for all $n$. Let $B\psi_n=b_n\psi_n$.
Let $\lambda_{m,n}=(\phi_m,\psi_n)$.
One calculates
\begin {eqnarray}
& &\Tr(\alpha{\bf 1}+H)(f(A)-f(B)-(A-B)f'(B))\cr
&=&\sum_n(\alpha+\epsilon_n)\sum_m|\lambda_{m,n}|^2
(f(a_n)-f(b_m)-(a_n-b_m)f'(b_m))\cr
&\ge&0
\end {eqnarray}
because, due to convexity of $f$ and to the assumption that
$\alpha{\bf 1}+H\ge 0$, each term in the previous sum
is non-negative.
\end {proof}

\begin {proofprop}
Let
\begin {equation}
f={q\over 2}f_2-f_q
\qquad\hbox{ with }
f_q(x)={x-x^q\over q-1}
\label {fdef}
\end {equation}
It is easy to check that $f$ is convex on the interval $[0,1]$
for $0<q\le 2$, $q\not=1$.
From the previous lemma with $A=\rho$ and $B=\rho_\alpha$ there follows that
\begin {eqnarray}
& &
{q\over 2}\Tr(\alpha{\bf 1}+H)
((\rho-\rho^2)-(\rho_\alpha-\rho_\alpha^2)
-(\rho-\rho_\alpha)(1-2\rho_\alpha))\cr
&\ge&
{1\over q-1}\Tr(\alpha{\bf 1}+H)
((\rho-\rho^q)-(\rho_\alpha-\rho_\alpha^q)
-(\rho-\rho_\alpha)(1-q\rho_\alpha^{q-1}))
\end {eqnarray}
The expression simplifies to (using that $\rho_\alpha$ commutes with $H$)
\begin {eqnarray}
\Tr(\alpha{\bf 1}+H)(\rho^q-\rho_\alpha^q
+q(\rho_\alpha-\rho)\rho_\alpha^{q-1})
&\ge&
{1\over 2}q(q-1)\Tr(\alpha{\bf 1}+H)(\rho-\rho_\alpha)^2\cr
&=&{1\over 2}q(q-1)||\rho-\rho_\alpha||_\alpha^2
\end {eqnarray}
Note that from the definition of $\rho_\alpha$ follows that
\begin {equation}
\Tr(\alpha{\bf 1}+H)(\rho-\rho_\alpha)\rho_\alpha^{q-1}=0
\end {equation}
Hence the expression simplifies further to
\begin {equation}
\Tr(\alpha{\bf 1}+H)(\rho^q-\rho_\alpha^q)
\ge {1\over 2}q(q-1)||\rho-\rho_\alpha||_\alpha^2
\end {equation}
This can be written as (\ref {underbound})
provided ${\cal G}^\beta_q(\rho)$ and ${\cal G}^\beta_q(\rho_\alpha)$
are finite.
\end {proofprop}

\noindent
The proposition implies that $\rho_\alpha$, with $k_BT=\alpha(q-1)$, is
the unique minimum of ${\cal G}^\beta_q$. This is the basis to prove
Theorem \ref {theoremsol}.

\section {Proof of Theorem \ref {theoremsol}}

Let $\alpha=\alpha_q(\beta)$.
Let $\rho$ be any density matrix for which
${\cal U}_q(\rho)$ is finite. We have to show that
\begin {equation}
{\cal F}^\beta_q(\rho)\ge {\cal F}^\beta_q(\rho_{\alpha})
\label {ineq0}
\end {equation}
with equality if and only if $\rho=\rho_\alpha$.
First assume that there exists $\gamma$ such that
\begin {equation}
{\cal S}_q(\rho)={\cal S}_q(\rho_{\gamma})
\label {ass1}
\end {equation}
Then
\begin {equation}
{\cal F}^\beta_q(\rho)\ge {\cal F}^\beta_q(\rho_{\gamma})
\label {ineq1}
\end {equation}
because $\rho_{\gamma}$ minimizes $\Tr\rho^qH$ given
that the entropy equals ${\cal S}_q(\rho_{\gamma})$
(the latter follows from proposition \ref {convex}).
But note that
\begin {equation}
{\cal F}^\beta_q(\rho_{\gamma})\ge {\cal F}^\beta_q(\rho_{\alpha})
\label {ineq2}
\end {equation}
because $\alpha=\alpha_q(\beta)$.
Indeed, relation (\ref {tcond}) was precisely derived by variation of
the free energy ${\cal F}^\beta_q(\rho_{\alpha})$ w.r.t.~$\alpha$.
Hence,
(\ref {ineq1}) and (\ref {ineq2}) together prove that $\rho_\alpha$
minimizes ${\cal F}^\beta_q$.
Still assuming (\ref {ass1}), let us show uniqueness of the equilibrium
density matrix. If
\begin {equation}
{\cal F}^\beta_q(\rho)={\cal F}^\beta_q(\rho_{\alpha})
\label {eqassum}
\end {equation}
then (\ref {ineq2}) is an equality. By the uniqueness of
proposition \ref {uniqsolprop}, there follows that $\alpha=\gamma$
(indeed, the free energy ${\cal F}_q^\beta(\rho_\gamma)$
is strictly decreasing for $\gamma<\alpha$ and strictly
increasing for $\gamma>\alpha$).
Hence, $\rho$ and $\rho_\alpha$ have the same entropy. But then,
equality (\ref {eqassum}) implies that they have also the same
energy. Now use that $\rho_\alpha$ is the unique density
matrix minimizing ${\cal G}^{\beta'}_q$ with $\beta'= 1/\alpha(q-1)$
(see proposition \ref {convex}).
Since also $\rho$ minimizes this expression
(it has the same value of $\Tr\rho^q H$ and of $\Tr\rho^q$)
there follows that $\rho=\rho_\alpha$.

Next assume that no $\gamma$ exists for which (\ref {ass1}) holds.
Consider first the case that
\begin {equation}
{\cal S}_q(\rho)>{\cal S}(\rho_\gamma)
\qquad \hbox{ for all } \gamma>-\epsilon_0
\label {slarger}
\end {equation}
and $\cal H$ is finite dimensional.
Then one has ${\cal S}_q(\rho)={\cal S}_q(\rho_\infty)$
with $\rho_\infty=(1/N){\bf 1}=\lim_{\gamma\rightarrow\infty}
\rho_\gamma$
($N$ is the dimension of $\cal H$).
A short calculation shows that for large $\gamma$ one has
\begin {equation}
{\cal U}_q(\rho_\gamma)={1\over N}\Tr H
-{q\over q-1}{1\over\gamma}
\left({1\over N}\Tr H^2-\left({1\over N}\Tr H\right)^2\right)
+\hbox {O}(\gamma^{-2})
\end {equation}
and
\begin {equation}
{\cal S}_q(\rho_\gamma)=k_B{1\over q-1}(1-N^{1-q})
+\hbox {O}(\gamma^{-2})
\end {equation}
Note that
\begin {equation}
{1\over N}\Tr H^2-\left({1\over N}\Tr H\right)^2>0
\end {equation}
because $H$ is not a multiple of the identity.
Hence for large $\gamma$ the function ${\cal F}^\beta_q(\rho_\gamma)$
is strictly increasing. This implies that
${\cal F}^\beta_q(\rho)>{\cal F}^\beta_q(\rho_\alpha)$.

If $\cal H$ is infinite dimensional then the strict inequality
${\cal S}_q(\rho)<k_B/(q-1)=
\lim_{\gamma\rightarrow\infty}{\cal S}_q(\rho_\gamma)$ holds for all $\rho$.
Hence (\ref {slarger}) cannot occur.
Remains the case that
\begin {equation}
{\cal S}_q(\rho)<{\cal S}(\rho_\gamma)
\qquad \hbox { for all }\gamma>-\epsilon_0
\label {assum2}
\end {equation}
Because entropy is an increasing function of $\alpha$
(proposition \ref {strictdecr}) it suffices now to look to
the limit $\alpha\downarrow-\epsilon_0$. In this limit
$\rho_\alpha$ converges to $\rho_g\equiv(1/m)E$ with $m$
the degeneracy of the ground state and $E$ the orthogonal
projection onto the ground state eigenvectors.
By assumption,
\begin {equation}
{\cal S}_q(\rho)\le {\cal S}_q(\rho_g)
=\lim_{\alpha\downarrow\epsilon_0}{\cal S}_q(\rho_\alpha)
\end {equation}
while necessarily ${\cal U}_q(\rho)\ge {\cal U}(\rho_g)=\epsilon_0$.
Hence one has
\begin {equation}
{\cal F}^\beta_q(\rho)\ge {\cal F}^\beta_q(\rho_g)
\label {ineq3}
\end {equation}
The inequality
\begin {equation}
{\cal F}^\beta_q(\rho_g)\ge {\cal F}^\beta_q(\rho_{\alpha})
\label {ineq4}
\end {equation}
follows because (\ref {ineq2}) holds for all $\gamma$, and hence also
for $\gamma\downarrow-\epsilon_0$.
Combination of (\ref {ineq3}) and (\ref {ineq4}) yields the
desired result.

Finally, from the analysis in proposition (\ref {uniqsolprop}) follows
that for $\gamma$ in $(-\epsilon_0, \alpha)$ the free energy
${\cal F}_q^\beta(\rho_\gamma)$ is a strictly decreasing function
of $\gamma$. Hence (\ref {ineq4}) is a strict inequality.
Therefore, unicity of the minimum follows also in this case.

\section {$0<q<1$-statistics}

Heuristic arguments lead to the conclusion that the equilibrium
density matrix, if it exists, 
is of the form $\rho'_\alpha$ given by
\begin {equation}
\rho'_\alpha={1\over\zeta'_\alpha}
\left[\alpha{\bf 1}- H\right]_+^{1/(1-q)}
\label {eqless}
\end {equation}
and
\begin {equation}
\zeta'_\alpha=\Tr\left[\alpha{\bf 1}- H\right]_+^{1/(1-q)}
\end {equation}
Here, $[A]_+$ is the restriction of $A$ to its positive part.
For self-adjoint $A$ with discrete spectrum this means that
$A\psi=\lambda\psi$ with $\lambda\in {\bf R}$
implies that $[A]_+\psi=[\lambda]_+\psi$,
with $[\lambda]_+=\max\{0,\lambda\}$.
The presence of $[\cdot]_+$ in (\ref {eqless}) is a high-energy cutoff
which is necessary to assure that
$\rho'_\alpha\ge 0$. Its presence complicates analytical calculations.
On the other hand, the operator $\left[\alpha{\bf 1}- H\right]_+$
is finite rank. Hence the energy ${\cal U}_q(\rho'_\alpha)$
exists for all $\alpha>\epsilon_0$. Let $H_\alpha$ denote minus the
negative part of $\alpha{\bf 1}- H$, i.e.
\begin {equation}
\alpha{\bf 1}- H=\left[\alpha{\bf 1}- H\right]_+-H_\alpha
\end {equation}
Then, using $\Tr {\rho'_\alpha}^qH_\alpha=0$, one calculates
\begin {eqnarray}
{\cal U}_q(\rho'_\alpha)
&=&{\Tr {\rho'_\alpha}^qH\over \Tr {\rho'_\alpha}^q}\cr
&=&\alpha-{\Tr {\rho'_\alpha}^q \left[\alpha{\bf 1}- H\right]_+
\over \Tr {\rho'_\alpha}^q}\cr
&=&\alpha-{{\zeta'_\alpha}^{1-q}\over\Tr{\rho'_\alpha}^q}
\label {energdecomp}
\end {eqnarray}
The expression for the free energy becomes
\begin {equation}
{\cal F}^\beta_q(\rho'_\alpha)=\alpha
-{{\zeta'_\alpha}^{1-q}\over\Tr{\rho'_\alpha}^q}
+{1\over\beta(1-q)}(1-\Tr{\rho'_\alpha}^q)
\end {equation}
Variation w.r.t.~$\alpha$ (assuming $\alpha\not=\epsilon_n$
for all $n$) gives
\begin {equation}
{\partial\ \over\partial\alpha}
{\cal F}^\beta_q(\rho'_\alpha)=
1-(1-q){{\zeta'_\alpha}^{-q}\over\Tr{\rho'_\alpha}^q}
{\partial\ \over\partial\alpha}\zeta'_\alpha
+\left({{\zeta'_\alpha}^{1-q}\over(\Tr{\rho'_\alpha}^q)^2}
-{1\over\beta(1-q)}\right)
{\partial\ \over\partial\alpha}\Tr{\rho'_\alpha}^q
\end {equation}
Using
\begin {equation}
{\partial\ \over\partial\alpha}\zeta'_\alpha=
{1\over 1-q}{\zeta'_\alpha}^q\Tr{\rho'_\alpha}^q
\label {dzetadalfa2}
\end {equation}
there follows
\begin {equation}
{\partial\ \over\partial\alpha}
{\cal F}^\beta_q(\rho'_\alpha)=
{1\over 1-q}\left({1\over \beta'_q(\alpha)}-{1\over\beta(1-q)}\right)
{\partial\ \over\partial\alpha}\Tr{\rho'_\alpha}^q
\label {freeder}
\end {equation}
with
\begin {equation}
\beta'_q(\alpha)={1\over 1-q}{(\Tr{\rho'_\alpha}^q)^2
\over {\zeta'_\alpha}^{1-q}}
\end {equation}

\begin {prop}
\label {strictincr}
$\Tr{\rho'_\alpha}^q$ is a non-decreasing function of $\alpha$.
If  $[\alpha{\bf 1}-H]_+$ is not a multiple of a projection operator
then $\Tr{\rho'_\alpha}^q$ is strictly increasing.
\end {prop}

\begin {proof}
The proof is very analogous to that of proposition (\ref {strictdecr}).
Without restriction assume that $\alpha\not=\epsilon_n$ for all $n$.
One has
\begin {equation}
{\partial\ \over\partial\alpha}\Tr{\rho'_\alpha}^q
={q\over 1-q}{\zeta'_\alpha}^{-1-q}
\left(f_\alpha(1)f_\alpha(2q-1)-f_\alpha(q)^2\right)
\label {eqb}
\end {equation}
with
\begin {equation}
f_\alpha(x)=\Tr[\alpha{\bf 1}-H]_+^{x/(1-q)}
\label {fdef2}
\end {equation}
The function $f_\alpha$ is log-convex (see the appendix).
It is strictly log-convex when $[\alpha{\bf 1}-H]_+$
is not a multiple of a projection operator.
Hence the r.h.s.~of (\ref {eqb}) is non-negative
resp.~strictly positive.
\end {proof}

One concludes that the derivative
of the free energy w.r.t.~$\alpha$ can vanish only if the
equation
\begin {equation}
\beta=\beta'_q(\alpha)
\label {tcond2}
\end {equation}
 is satisfied
(assuming that $\alpha$ is large enough so that $[\alpha{\bf 1}-H]_+$
is not a multiple of a projection operator).
Note that $\beta'_q(\alpha)$ can be written out as
\begin {equation}
\beta'_q(\alpha)={1\over 1-q}
{
\left(\Tr[\alpha{\bf 1}-H]_+^{q/(1-q)}\right)^2
\over
\left(\Tr[\alpha{\bf 1}-H]_+^{1/(1-q)}\right)^{1+q}
}
\end {equation}

\section {Thermodynamic stability for $q<1$}

Up to here the analogy between $q<1$ and $q>1$ is almost complete. In
particular, $\beta'_q(\alpha)$ differs from $\beta_q(\alpha)$ by the
factor $1/(q-1)$, which is replaced by $1/(1-q)$, and by replacing
$\alpha{\bf 1}+ H$ by $[\alpha{\bf 1}-H]_+$. However, $\beta_q(\alpha)$
is a strictly decreasing function of $\alpha$, with range $(0,+\infty)$
(proposition \ref {uniqsolprop}). It is in general {\sl not} possible to
prove this statement for $\beta'_q(\alpha)$. In addition, extremes of
${\cal F}^\beta_q(\rho'_\alpha)$ can occur at
$\alpha=\epsilon_n,n=0,1,\cdots$ where the derivative of the free energy
may not exist if $q\le 1/2$.

In fact, further aspects of the thermodynamic formalism may go wrong. It
can happen that the map $\alpha\rightarrow {\cal
F}^\beta_q(\rho'_\alpha)$ is not bounded below. In such a case no
equilibrium state can exist. It is obvious, given an infinite
dimensional Hilbert space $\cal H$, to expect that ${\cal
U}_q(\rho'_\alpha)$ increases linearly in $\alpha$. A necessary
condition for thermodynamic stability is then that ${\cal
S}_q(\rho'_\alpha)$ increases slower than $\alpha$. This is the subject
of the next proposition.

\begin {prop}
Let $0<q<1$ and assume that $q+q_c>2$. Then there exists $\lambda<1$
and a constant $K$ such that
\begin {equation}
{\cal S}_q(\rho'_\alpha)\le K(\alpha-\epsilon_0)^{(1-q)/(q_c-1)},
\qquad \alpha>\epsilon_0
\label {sla}
\end {equation}
\end {prop}

\begin {proof}
One has, using notation (\ref {fdef2}),
\begin {equation}
{\cal S}_q(\rho'_\alpha)
=k_B{1\over 1-q}
\left({f_\alpha(q)
\over f_\alpha(1)^q}
-1\right)
\label {sexpr}
\end {equation}
Because $f_\alpha$ is log-convex one has
\begin {equation}
f_\alpha(q)\le f_\alpha(1)^qf_\alpha(0)^{1-q}
\end {equation}
But $f_\alpha(0)$ equals the number of eigenvalues of $H$ strictly
less than $\alpha$. From the definition of $q_c$
follows that $\gamma$ exists such that
\begin {equation}
\epsilon_n-\epsilon_0\ge\gamma n^{q_c-1}
\end {equation}
holds for all $n$.
Hence one has
\begin {equation}
f_\alpha(0) \le \left({\alpha-\epsilon_0\over\gamma}\right)^{1/(q_c-1)}
\label {termcntub}
\end {equation}
One obtains 
\begin {equation}
{\cal S}_q(\rho'_\alpha)
\le k_B{1\over 1-q}
f_\alpha(0)^{1-q}
\le k_B{1\over 1-q}
\left({\alpha-\epsilon_0\over\gamma}\right)^{(1-q)/(q_c-1)}
\end {equation}
This proves (\ref {sla}).
\end {proof}

No conditions will be given to assure that ${\cal U}_q(\rho'_\alpha)$
increases linearly with $\alpha$. Indeed, less is needed because it will
be assumed that ${\cal S}_q(\rho'_\alpha)$ increases as $\alpha^\kappa$
with $\kappa<1$. Let us start by showing that it is not automatically
the case that ${\cal U}_q(\rho'_\alpha)$ increases linearly with
$\alpha$. The following result states that the energy is at most the
average value of the occupied energy levels, which is obvious because
low energy levels have higher occupancy than high energy levels.

\begin {prop}
One has for all $\alpha>\epsilon_0$
\begin {equation}
{\cal U}_q(\rho'_\alpha)
\le{1\over N}\sum_{n=0}^{N-1}\epsilon_n
\label {energupb}
\end {equation}
with $N$ the number of eigenvalues $\epsilon_n$ satisfying
$\epsilon_n<\alpha$.
\end {prop}

\begin {proof}
From (\ref {energdecomp}) follows that
\begin {equation}
\alpha-{\cal U}_q(\rho'_\alpha)
={f_\alpha(1)\over f_\alpha(q)}
=\alpha{
\sum_n[1-{\epsilon_n/\alpha}]_+^{1/(1-q)}
\over
\sum_n[1-{\epsilon_n/\alpha}]_+^{q/(1-q)}
}
\label {lln1}
\end {equation}
Now, for any sequence of positive numbers $(\lambda_n)_n$ the
function
\begin {equation}
x\rightarrow {\sum_n\lambda_n^{x+1}\over \sum_n\lambda_n^{x}}
\end {equation}
is increasing. To see this, take the derivative w.r.t.~$x$
and use that $\lambda$ and $\log\lambda$ are positively
correlated. Hence, (\ref {lln1}) can be estimated by
\begin {eqnarray}
\alpha-{\cal U}_q(\rho'_\alpha)
&\ge&
\alpha{
\sum_n[1-{\epsilon_n/\alpha}]_+
\over
\sum_n[1-{\epsilon_n/\alpha}]_+^0
}\cr
&=&\alpha-{1\over N}\sum_{n=0}^{N-1}\epsilon_n
\label {lln2}
\end {eqnarray}
\end {proof}

The proposition shows that, if one wants that ${\cal U}_q(\rho'_\alpha)$
increases linearly in $\alpha$ then at least
$(1/N)\sum_{n=0}^{N-1}\epsilon_n$
should increase linearly in $\epsilon_N$. It is easy to produce an example
which does not satisfy this requirement. Let $\epsilon_n=a^n$
with $a>1$. Then one calculates that
\begin {equation}
{1\over N}\sum_{n=0}^{N-1}\epsilon_n={1\over N}{\epsilon_N-1\over a-1}
\end {equation}
which increases slower that linearly in $\epsilon_N$.

Let $m$ denote the multiplicity
of the ground state energy $\epsilon_0$. Then 
$\epsilon_m$ is the energy of the first excited state.
One has
\begin {equation}
\beta'_q(\epsilon_m)={1\over 1-q}{m^{1-q}\over \epsilon_m-\epsilon_0}
\label {betaplus}
\end {equation}

\begin {prop}
Assume that $2<q_c+q$. Assume also that $a>1$ and $N_0$ exist
such that
\begin {equation}
\epsilon_N\ge a{1\over N}\sum_{n=0}^{N-1}\epsilon_n,
\qquad \hbox{ for all }N\ge N_0
\label {asonav}
\end {equation}
Then the range of $\beta'_q$ is $(0,+\infty)$.
\end {prop}

\begin {proof}
Let us start by proving that
\begin {equation}
\lim_{\alpha\rightarrow\infty}\beta'_q(\alpha)=0
\end {equation}
Using
\begin {equation}
f_\alpha(q)^2\le f_\alpha(1)^qf_\alpha(0)^{1-q}
\end {equation}
there follows
\begin {equation}
\beta'_q(\alpha)={f_\alpha(q)^2\over f_\alpha(1)^{1+q}}
\le \left({f_\alpha(0)^2\over f_\alpha(1)}\right)^{1-q}
\end {equation}
Using 
\begin {equation}
f_\alpha(1-q)\le f_\alpha(1)^{1-q}f_\alpha(0)^{q}
\end {equation}
the latter becomes
\begin {equation}
\beta'_q(\alpha)\le {f_\alpha(0)^{2-q}\over f_\alpha(1-q)}
\label {leqn1}
\end {equation}
Now note that, using assumption (\ref {asonav}),
one has for $N$ large enough
\begin {eqnarray}
{1\over N}\sum_{n=0}^{N-1}\epsilon_n
&=&{N-1\over N}{1\over N-1}\sum_{n=0}^{N-2}\epsilon_n
+{1\over N}\,\epsilon_{N-1}\cr
&\le&\left({N-1\over N}{1\over a}
+{1\over N}\right)
\epsilon_{N-1}
\label {navinc}
\end {eqnarray}
There follows
\begin {eqnarray}
{f_\alpha(1-q)\over f_\alpha(0)}
&=&\alpha{1\over N}\sum_{n=0}^{N-1}(1-{\epsilon_n\over\alpha})\cr
&=&\alpha-{1\over N}\sum_{n=0}^{N-1}\epsilon_n\cr
&\ge& \alpha-\left({N-1\over N}{1\over a}
+{1\over N}\right)
\epsilon_{N-1}\cr
&\ge&\alpha{N-1\over N}\left(1-{1\over a}\right)
\end {eqnarray}
(as before, $N$ is the number of eigenvalues less than $\alpha$).
Using this result and (\ref {termcntub}), (\ref {leqn1}) can be written as
\begin {equation}
\beta'_q(\alpha)\le
{1\over\alpha}{2a\over a-1}f_\alpha(0)^{1-q}
\le {1\over\alpha}{2a\over a-1}
\left({\alpha-\epsilon_0\over\gamma}\right)^{(1-q)/(q_c-1)}
\end {equation}
The latter tends to zero because of the assumption that
$2<q+q_c$.

Next consider the limit $\alpha\rightarrow\epsilon_0$. One has
\begin {equation}
\beta'_q(\alpha)={m^{1-q}\over \alpha-\epsilon_0},
\qquad \alpha\in (\epsilon_0,\epsilon_m]
\end {equation}
It takes on any value in the interval
$[(1-q)\beta'_q(\epsilon_m),+\infty)$. Now, because $\beta'_q(\alpha)$
is a continuous function of $\alpha$ it takes on all values in the
interval $(0,+\infty)$.
\end {proof}

Note that condition (\ref {asonav}) implies that the average of
eigenvalues $\epsilon_0$ to $\epsilon_{N-1}$ is a strictly increasing
function of $N$. This statement is weaker than the condition that
it should increase linearly in $\epsilon_N$, but suffices for our
purposes. An example of a spectrum which does not satisfy (\ref {asonav})
is given by $\epsilon_0=1$ and
\begin {equation}
\epsilon_N=(n+1)!
\qquad\hbox{ for }N\in \{n!,\cdots,(n+1)!-1\}\hbox { and }n>0
\end {equation}
The eigenvalue $n!$ has degeneracy $(n-1)\times(n-1)!$. The average
of the first $N!$ terms equals
\begin {equation}
N!\left(1-\sum_{n=1}^{N-1}n\left({n!\over N!}\right)^2\right)
\ge N!(1-1/N)
\end {equation}
Hence(\ref {navinc}) does not hold for this example.

\begin {prop}
\label {uniqsolprop00}
Under the conditions of the previous proposition,
the map $\alpha\rightarrow{\cal F}_q^\beta(\rho'_\alpha)$
has at least one absolute minimum for any $\beta>0$.

Assume that $2<q_c+q$ and that condition (\ref {asonav}) is satisfied.
For any $\beta>0$ the equation $\beta=\beta'_q(\alpha)$ has at least
one solution. For $\beta\ge\beta'_q(\epsilon_m)$ 
the ground state of the system is a solution.
For any  .
\end {prop}

\begin {proof}
Because $\beta'_q(\alpha)$
tends to zero as $\alpha$ tends to infinity, it follows that for large
$\alpha$ both factors of (\ref {freeder}) are strictly positive. Hence
the free energy is strictly increasing for large $\alpha$. Since it is a
continuous function, piecewise differentiable, and bounded from below by
some function linear in $\alpha$, it has at least one absolute minimum.
\end {proof}

Another feature of $q<1$-thermodynamics is the non-uniqueness of
density matrices minimizing the free energy. The following example shows
that phase transitions can occur even in systems with finitely many
degrees of freedom as considered here.

\begin {example}
Let the hamiltonian be given by the 2-by-2 matrix
\begin {equation}
H=\left(\matrix{-\mu &0\cr 0 & \mu}\right)
\end {equation}
with $\mu>0$. A short calculation shows that
\begin {eqnarray}
{\cal F}_{1/2}^\beta(\rho'_\alpha)
&=& -\mu,\qquad\qquad \hbox{ if } -\mu<\alpha\le\mu\cr
&=&-\mu{1-\kappa\over 1+\kappa}
+{2\over\beta}\left(1-{1+\kappa\over\sqrt{1+\kappa^2}}\right),
\qquad \hbox{ if }\alpha\ge\mu
\end {eqnarray}
with
\begin {equation}
\kappa={\alpha-\mu\over \alpha+\mu}
\end {equation}
For $\beta\mu\le 1$ the free energy ${\cal F}^\beta_{1/2}$
has a unique minimum at some value
of $\alpha>\mu$. In a small interval $\beta\mu\in (1,1+\epsilon)$,
it has a relative minimum for $\alpha\in [-\mu,\mu]$ and an absolute
minimum at some value of $\alpha>\mu$. Finally, for $\beta\mu>1+\epsilon$,
the ground state (corresponding with $\alpha\in [-\mu,\mu]$)
is the absolute minimum.
This means that the transition to the ground state
occurs at finite temperature and is a phase transition of first order.
\end {example}

\section {High-energy cutoff}

The existence of thermodynamic equilibrium has been discussed in the
previous section. Here, existence of a unique equilibrium state is
assumed. It is shown that it is necessarily
of the form $\rho'_\alpha$ with $\alpha$ a solution of (\ref {tcond2}).
Hence, a special feature of $q<1$-statistics is that the
equilibrium density matrix is a finite rank operator.
This means that the high energy levels
of $H$ are not occupied. In particular,
for low enough temperatures ($\beta\ge \beta'_q(\epsilon_m)$)
the equilibrium density matrix is $E/m$, i.e.~only the ground state
is occupied. 

\begin {theorem}
\label {qlessone}
Let $0<q<1$. Let $H$ be a self-adjoint operator of the
Hilbert space ${\cal H}$.
Assume that either
\begin {itemize}
\item ${\cal H}$ is finite dimensional and $H$ is not a multiple of
$\bf 1$
\end {itemize}
\noindent
or
\begin {itemize}
\item ${\cal H}$ is infinite dimensional,
the spectrum of $H$ is discrete, bounded from below,
with isolated eigenvalues of finite multiplicity.
\end  {itemize}
Let $\beta > 0$. Assume that
the map $\alpha\rightarrow {\cal F}^\beta_q(\rho'_\alpha)$,
defined on the interval $(\epsilon_0,+\infty)$,
has a unique minimum at a finite value $\alpha_m$ of $\alpha$.
Then the free energy $\rho\rightarrow {\cal F}^\beta_q(\rho)$
has a unique minimum. It occurs at $\rho=\rho'_{\alpha_m}$.
\end {theorem}

\noindent
The proof of the theorem follows now.

Let $\Tr_\alpha$ denote the partial trace over the subspace of $\cal H$
spanned by the eigenvectors $\psi_n$ for which
$\alpha-\epsilon_n>0$. Introduce a semi-norm defined by
\begin {equation}
||A||_\alpha^2=\Tr_\alpha (\alpha {\bf 1}-H)|A|^2
=\Tr[\alpha {\bf 1}-H]_+|A|^2
=\sum_n[\alpha-\epsilon_n]_+||A\psi_n||^2
\end {equation}

Let ${\cal G}^\beta_q(\rho)$ be given by (\ref {oldfreenerg}).

\begin {prop}
\label {convex2}
Let $0<q<1$. Let $\alpha=k_BT/(1-q)$
and assume that $\alpha>\epsilon_0$.
Then one has
\begin {equation}
{\cal G}^\beta_q(\rho)-{\cal G}^\beta_q(\rho'_\alpha)
\ge {1\over 2}q(1-q)\,||\rho-\rho'_\alpha||_\alpha^2
+\Tr \rho^qH_\alpha +q{\zeta'_\alpha}^{1-q}(1-\Tr_\alpha\rho)
\label {underbound2}
\end {equation}
for any density matrix $\rho$.
\end {prop}

\begin {proof}
The proof is analogous to that of proposition \ref {convex}.
From Klein's inequality,
as given by lemma \ref {klein}, but with $\alpha{\bf 1}+H$ replaced by
$[\alpha{\bf 1}-H)]_+$, one obtains
\begin {eqnarray}
& &{q\over 2}\Tr_\alpha [\alpha{\bf 1}-H)]_+
\times
\bigg((\rho-\rho^2)-({\rho'_\alpha}-{\rho'_\alpha}^2)
-(\rho-\rho'_\alpha)(1-2\rho'_\alpha)\bigg)\cr
&\ge&-{1\over 1-q}\Tr_\alpha [\alpha{\bf 1}-H)]_+\cr
& &\times
\bigg((\rho-\rho^q)-(\rho'_\alpha-{\rho'_\alpha}^q)
 -(\rho-\rho'_\alpha)(1-q{\rho'_\alpha}^{q-1})\bigg)
\end {eqnarray}
The expression simplifies to
\begin {equation}
-{q\over 2}||\rho-\rho'_\alpha||_\alpha^2
\ge -{1\over 1-q}\Tr_\alpha [\alpha{\bf 1}-H)]_+
\times
\left(-\rho^q+{\rho'_\alpha}^q+q(\rho-\rho'_\alpha){\rho'_\alpha}^{q-1}
\right)
\end {equation}
Using the definition of $\rho'_\alpha$ one shows that
\begin {equation}
\Tr_\alpha [\alpha{\bf 1}-H)]_+(\rho-\rho'_\alpha){\rho'_\alpha}^{q-1}
={\zeta'_\alpha}^{1-q}(\Tr_\alpha\rho-1)
\end {equation}
and
\begin {equation}
{\cal G}^\beta_q(\rho)-{\cal G}^\beta_q(\rho'_\alpha)
=-\Tr_\alpha [\alpha{\bf 1}-H)]_+
\left(\rho^q-{\rho'_\alpha}^q\right)+\Tr\rho^q H_\alpha
\end {equation}
Putting the pieces together yields (\ref {underbound2}).
\end {proof}

The proposition shows that $\rho'_\alpha$ is the unique
minimum of ${\cal G}^\beta_q(\rho)$ (note that $H_\alpha\ge 0$
and $\Tr_\alpha\rho\le 1$).

Let $\rho$ be any density matrix for which
${\cal U}_q(\rho)$ and ${\cal S}_q(\rho)$ are finite.
We have to show that
\begin {equation}
{\cal F}^\beta_q(\rho)\ge {\cal F}^\beta_q(\rho'_{\alpha_m})
\label {ineq00}
\end {equation}
First consider the case that $\gamma>\epsilon_0$ exists such that
${\cal S}_q(\rho)={\cal S}_q(\rho'_\gamma)$. 
Then, by proposition \ref {convex2},
\begin {equation}
{\cal U}_q(\rho)\ge {\cal U}_q(\rho'_\gamma)
\qquad
\hbox{ and hence }
{\cal F}^\beta_q(\rho)\ge {\cal F}^\beta_q(\rho'_\gamma)
\label {ineq001}
\end {equation}
By the assumption made in the formulation of the theorem one has
\begin {equation}
{\cal F}^\beta_q(\rho'_\gamma)\ge {\cal F}^\beta_q(\rho'_{\alpha_m})
\label {ineq002}
\end {equation}
Combination of both inequalities yields (\ref {ineq00}).
If equality holds in
(\ref {ineq00}), then it holds also in (\ref {ineq001}) and (\ref {ineq002}),
and implies that $\rho'_{\alpha_m}=\rho'_\gamma=\rho$.

Before going on let us prove the following.

\begin {lemma}
Assume that $\cal H$ is infinite dimensional. Then one has
\begin {equation}
\lim_{\alpha\rightarrow\infty}{\cal S}_q(\rho'_\alpha)=+\infty
\end {equation}
\end {lemma}

\begin {proof}
One has
\begin {eqnarray}
{f_\alpha(q)\over f_\alpha(1)^q}
&=&{\sum_n\left[1-\epsilon_n/\alpha\right]_+^{q/(1-q)}
\over \left(\left[1-\epsilon_n/\alpha\right]_+^{1/(1-q)}\right)^q}\cr
&\ge& \left(\left[1-\epsilon_n/\alpha\right]_+^{q/(1-q)}\right)^{1-q}
\label {leqn2}
\end {eqnarray}
because term by term holds
\begin {equation}
\left[1-\epsilon_n/\alpha\right]_+^{q/(1-q)}
\ge \left[1-\epsilon_n/\alpha\right]_+^{1/(1-q)}
\end {equation}
Now, the number of terms in (\ref {leqn2}) tends to infinity
while each of the terms tends to 1. Hence it is clear that
(\ref {leqn2}) tends to infinity. Since ${\cal S}(\rho'_\alpha)$
is proportional to $f_\alpha(q)/f_\alpha(1)^q$ (see (\ref {sexpr}))
the lemma is proved.
\end {proof}

Next assume that no $\gamma>\epsilon_0$ exists for which 
${\cal S}_q(\rho)={\cal S}_q(\rho'_\gamma)$ holds.
there are two possibilities.
First assume that
\begin {equation}
{\cal S}_q(\rho)=\lim_{\gamma\rightarrow+\infty}{\cal S}_q(\rho'_\gamma)
<+\infty
\end {equation}
Then, necessarily by the previous lemma, $\cal H$ is $N$-dimensional
and $\rho={\bf 1}/N$.
For large $\gamma$ a straightforward calculation shows that
\begin {equation}
{\cal U}_q(\rho'_\gamma)
={1\over N}\Tr H -{1\over\gamma}{q\over 1-q}
\left({1\over N}\Tr H^2-\left({1\over N}\Tr H\right)^2\right)
+\hbox{O}(\gamma^{-2})
\end {equation}
and
\begin {equation}
{\cal S}_q(\rho'_\gamma)
=k_B{1\over 1-q}(N^{1-q}-1)
+\hbox{O}(\gamma^{-2})
\end {equation}
This shows that ${\cal F}^\beta_q(\rho'_\gamma)$
is strictly increasing for large enough $\gamma$.
Indeed, by convexity, using that $H$ is not a multiple of the identity,
one shows that
\begin {equation}
{1\over N}\Tr H^2-\left({1\over N}\Tr H\right)^2>0
\end {equation}
One concludes therefore that
${\cal F}^\beta_q(\rho)>{\cal F}^\beta_q(\rho'_{\alpha_m})$.

Remains the case that ${\cal S}_q(\rho)<{\cal S}_q(\rho'_\gamma)$
for all $\gamma>\epsilon_0$.
For $\epsilon_0<\gamma\le\epsilon_m$ is $\rho'_\gamma=E/m$. Hence
${\cal U}_q(\rho'_\gamma)=\epsilon_0$. Because ${\cal U}_q(\rho)$
cannot be smaller than $\epsilon_0$ there follows that
\begin {equation}
{\cal U}_q(\rho)-T{\cal S}_q(\rho)>
{\cal U}_q(\rho'_\gamma)-T{\cal S}_q(\rho'_\gamma)
\qquad \hbox{ for }\gamma=\epsilon_m
\end {equation}
This implies (\ref {ineq00}).

\section {Summary and discussion}

This paper studies the canonical ensemble of non-extensive thermodynamics for 
quantum mechanical systems with a finite number of degrees of freedom. Two 
different situations occur depending on whether the entropic parameter $q$ is 
larger than 1 or smaller than 1. If the Hilbert space is infinite dimensional 
then for $q>1$ existence of the trial density matrix requires that $q$ is less 
than some critical value $q_c$ which depends on $H$. Under the extra condition 
$q\le 2$ theorem 1 proves that the trial density matrix is the unique 
equilibrium density matrix.

For $q<1$ it can happen that the free energy is not bounded below so that no 
equilibrium density matrix can exist. To exclude this possibility the assumption 
$2<q_c+q$ has been made, as well as a further condition on the spectrum of $H$. 
In addition, even if the free energy is bounded below, it is possible that the 
minimum of the free energy is non-unique. In other words, $q<1$-statistics can 
produce phase transitions even in systems with a finite number of degrees of 
freedom. As a consequence, a less general result than for $q>1$ is obtained. 
Proposition \ref {uniqsolprop00} proves that the free energy has at least one 
minimum in the set of trial density matrices. Theorem 2 proves that, if this 
minimum is unique, then the trial density matrix is also the equilibrium state 
of the system. Note that the trial density matrix is a finite rank operator. 
Hence, a special feature of $q<1$-statistics is that high energy levels are not 
occupied. This result supports the interpretation of $q<1$-statistics as the 
statistics of non-extensive systems, or of systems in equilibrium with a finite 
heath bath \cite {PP94}.

The high degree of stability of the $q>1$-theory finds its origin in the
fact that $q$-entropy is bounded for $q>1$. In case $q<1$ entropy can
diverge and the energy-entropy balance can become unstable, which is one
of the reasons why density matrices can exist with arbitrary small free
energy. The other factor favoring thermodynamic instability is the
normalization of the energy functional. The denominator $\Tr\rho^q$ in
(\ref {energ}) keeps the energy small while entropy increases.

The situation for $q>2$ has not been considered for technical reasons
(the basic convexity estimates rely on a comparison between
$q<2$-statistics with $q=2$-statistics). It is not yet clear how to
tackle the $q>2$-case.

\begin {appendix}
\section* {Appendix}

The following result is well-known.

\begin {lemma}
Let $K$ be a self-adjoint operator such that $\exp(-x K)$ is
trace-class for all $x>0$. Then the function
\begin {equation}
f(x)=\Tr e^{-x K}
\end {equation}
is log-convex.
If $K$ is not a multiple of the identity ${\bf 1}$
then $f$ is strictly log-convex.
\end {lemma}

\begin {proof}

One has
\begin {equation}
{\partial^2\ \over\partial x^2}\ln f(x)
={\Tr K^2e^{-x K}\over\Tr e^{-x K}}
-\left({\Tr Ke^{-x K}\over\Tr e^{-x K}}\right)^2
=\Tr\rho X^2\ge 0
\label {jens}
\end {equation}
with
\begin {equation}
\rho={e^{-x K}\over\Tr e^{-x K}}
\end {equation}
and $X=K-\Tr\rho K$.

Assume now that the rhs of (\ref {jens}) equals zero.
Then $X=0$ follows and hence $K=(\Tr K){\bf 1}$.
\end {proof}

Now write
\begin {equation}
(\alpha{\bf 1}-\beta(1-q)H)^{1/(1-q)}=e^{-K}
\end {equation}
Then
\begin {equation}
f_\alpha(x)=\Tr(\alpha{\bf 1}-\beta(1-q)H)^{x/(1-q)}=\Tr e^{-xK}
\end {equation}
is trace-ce-class for all $x>0$ by assumption,
so that the previous proposition can be applied
to obtain that $f_\alpha$ is log-convex.

On the other hand, if $f_\alpha$ is defined by
\begin {equation}
f_\alpha(x)=\Tr[\alpha{\bf 1}-H]_+^{x/(1-q)}
\end {equation}
then let
\begin {equation}
[\alpha{\bf 1}-H]_+^{1/(1-q)}=e^{-K}
\end {equation}
on the sub-Hilbert space spanned by the eigenvectors of
$[\alpha{\bf 1}-H]_+$ with strictly positive eigenvalue.
Application of the lemma leads then to log-convexity,
and strict log-convexity if $[\alpha{\bf 1}-H]_+$ is
not the identity operator of the sub-Hilbert space.

\end  {appendix}


\end {document}